\documentclass{amsart}

\usepackage{amssymb}
\usepackage{amsfonts}
\usepackage{amsmath}
\usepackage{graphicx}%
\usepackage[legalpaper,bookmarks=true,colorlinks=true,linkcolor=blue,citecolor=blue]{hyperref}
\usepackage{fancyhdr}
\usepackage{color}
\usepackage[mathlines]{lineno}
\usepackage{lscape}
\usepackage{epsfig}
\usepackage{float}
\usepackage{textcomp}
\usepackage[]{listings}

\theoremstyle{plain}

\numberwithin{equation}{section}

\begin{document}
\title[VB and R codes using Households databases available in the NSI's]{VB and R codes using Households databases available in the NSI's : A prelude to statistical applied studies}
\author{Gane Samb Lo $^{1,2,3}$}

\maketitle

\begin{abstract}
 We describe the main features of the households databases we can find in most of our National Statistics Institute. We provide algorithms aimed at extracting a diversity of variables on which different statistical procedures may be applied. Here, we particularly focus on the scaled income, as a beginning. Associated codes (MS Visual Basic \texttrademark \ and \textbf{R} codes) have been successfully tested and delivered in the text and in a separate file..\medskip
 
\noindent 
$^{1}$ LERSTAD, Gaston Berger University, Saint-Louis, S\'en\'egal (Gane Samb Lo, Modou Ngom). \\
$^{2}$ Affiliated to LSTA, Universit\'e Pierre et Marie Curie, Paris, France (Gane Samb LO)\\
$^{3}$ Associated to African University of Sciences and Technology, AUST, Abuja, Nigeria (Gane Samb Lo)\\
\newline

\noindent \textit{Corresponding author}. Gane Samb LO, gane-samb.lo@ugb.edu.sn\\

\bigskip  \noindent 
\textbf{keywords and phrases} : National Statistical Institutes (NSI); Preludes to Statistical Studies; Data storing;  Data handling; Households databases; Algorithms; computer programs and codes\\
\noindent {\textbf{AMS 2010 Classification subjects}}. 68P01; 68P05; 68N01
\end{abstract}

\Large

\section{Introduction}

\noindent In this paper, we want to share our experience in using some household databases one can find in most of African National Statistical institutes (\textit{NSI}).\\

\noindent In most of the African countries, a number of surveys have been and continue to be conducted at a national wide level. The
\begin{center}
\cite{asyb2016}
\end{center}
\noindent provided by the website of the African Center for Statistics \textit{ACS}, uses data from 54 African countries. So, at the continental level, the availability of such households databases amounts to a considerable number of interesting and useful data collections. At the same time, it is very strange and disappointing that only a very few number of research works has been undertaken by African Scholars, at least, scholars in the Mathematics and Statistics Departments.\\

\noindent As mentioned in \cite{gslo2014ajas}, I witnessed, as editor of \textit{Afrika Statistika} for almost a decade and as a member of a number of Master and PhD theses juries, that most of our colleagues and our students use data sets concerning western countries and picked out from the books to illustrate their theoretical works. I reviewed a very interesting \textit{PhD} thesis, from Central Africa, on Generalized Poisson Laws in which the candidate used data from England, exactly the number rain days in London. In my report, I pointed out such statistics do exist in meteorology stations and I recommended to contact them and to get the data.\\

\noindent The explanations to such an unacceptable situation, which I think is more dramatic in French speaking countries, are contradictory. While in the \textit{NSI}, we may hear that the African university scholars are hardly interested in applied works, and prefer to confine themselves in Theory, far from the real world. From the universities, it is sometimes said that the willingness to undertake applied works is real but the access to the data is not that simple.\\

\noindent Yet, there is a great number of studies using African data in the literature, mostly by international researchers. The works undertaken by the World Bank Living Standards Measurement Study (\textit{LSMS}) are enough to show that.\\

\noindent We simply advocate that all parties to conjugate their efforts to make emerge major research teams and research fields on African data and at the same time, to use extensively local and real data in the every day teaching activities in our universities.\\

\noindent In each country, a National Statistical National System (\textit{NSN}) is set to supervise national data production processes and its use. The universities are usually members of the \textit{NSN} Board and \textit{NSI}'s are the implementing bodies.\\

\noindent There must not be any problem to get the data when the request comes from the official heads of the Universities. This is the way used to get the data on which the current paper is based on.\\

\noindent This paper will be a report of data handling of households survey databases. We expect to be followed by colleagues in our statistics and and mathematics departments. In doing so, the \textit{AJAS} platform may become, very soon, a place where African data are explored and discussed at the light of most powerful and new theories and statistics trends.\\ 

\noindent For than a decade, we are working on statistical studies of Welfare issues using the highly elaborated theory of functional empirical processes. But, we used to apply the our findings on real data from Households databases from from Senegal, Mali and Mauritania  (See for example : \cite{lo6}, \cite{losallseck}, \cite{lo}, \cite{mergane1}, \cite{halo}, \cite{haidara-lo}).\\

\noindent Here, we want to introduce simple methods on handling the databases before undertaking statistical procedures. we hope that this may help and encourage others to acquire the databases and get involved in the study of such treasures. In this paper, we are going to do the following :\\

\noindent (a) Describe common features of these databases.\\

\noindent (b) Show how to extract pertinent information from these databases.\\

\noindent (c) Share useful and replicable \textit{R} and \textit{MS Visual Basic} codes.\\

\noindent We hope that this paper will inspire readers to get similar databases and to use, update or expand the given codes to prepare data and undertake applied statistics studies from them. At the end, sharing those studies in the pages of African Journal of Applied Statistics or elsewhere, may be an efficient way to achieving the awakening we are calling for.\\

\section{Households databases}

\noindent In the $80$'s, in the $90$'s and in the first decades of the 21th century, a number of surveys have been conducted. Nowadays, surveys destined to collect longitudinal data are available. But this paper will be devoted to households surveys with the objectives of showing how researchers may be quickly introduced to be ready use such data.\\

\subsection{Description of Households data} $ $\\   
 
\noindent \textbf{A - Organization of the files}.\\

\noindent The households databases (\textit{HDB} in short) gather information collected in a sample of households. In each household, all the member respond to the questions. At the arrival, a \textit{HDB} usually contains three main files.\\

\noindent (1) A person file which contains the responses of individuals of all households. This file contains the disaggregated data.\\

\noindent (2) A household file which contains aggregated data at the household level provided experts of the \textit{NSI}'s.\\

\noindent (3) A meta-file or  dictionary which explains the features of each variable.\\

\noindent These files are provided in \textit{MS Excel}\texttrademark \ or in \textit{IBM SPSS}\texttrademark \ in most of the \textit{NSI}'s. When 
\textbf{IBM SPSS}\texttrademark \ is used, the person file is given in an \textit{SPSS}\texttrademark \ fold contained the data and the meta-file.\\

\noindent Figure \ref{fig1} shows how the data are displayed.\\

\begin{figure}[htbp]
	\centering
	\caption{A view at the data file in a household database}
	\label{fig1}
\end{figure}

\noindent Figure \ref{fig2} provides information of the variables, which are listed in Column 2 called \textbf{Name}. For each variable, the type (categorical, logical, numerical, etc. is precised). For a factor or label variable, the range of the values are given in Column \textbf{Values}.\\

\begin{figure}[htbp]
	\centering
	\caption{A view at the dictionary metafile in a household database}
	\label{fig2}
\end{figure}

\noindent For MS Excel \texttrademark \ files, a separate dictionary files is given in replacement of the \textit{IBM SPSS} \texttrademark \  meta-file.\\

\noindent It is not excluded that different software packages or other file formats (like \textit{CSV}) may be used in some \textit{NSI}'s. However, the files organization must be similar and automatic conversions are available.

\bigskip \noindent \textbf{B - Organization of the data}.\\

\noindent (a) First of all, the country is organized into geographical regions, districts or areas. The nature of these areas depends on each country. In Senegal, for example, they are the administrative \textbf{regions}. From one survey to another, the administration division system may change.\\

\noindent Here, the variable indicating this area in the person file is denoted by \textbf{Region}.\\

\noindent (b) Next, each region is divided into a number of districts. The variable factor associated to this district is named \textbf{District} or \textbf{Milieu}. This variable is re-initiated to one for each region.\\

\noindent (c) In each district, a number of households are included in the survey. The factor variable, in some \textbf{HDB}'s, is not re-initiated to zero. Instead, the numbering continues and the last value is the total number of households included in the survey. It is named \textbf{Household} or \textbf{Menage}\\

\noindent The computer programs we will give handle both situations where the numbering is re-initiated or not.\\

\noindent (d) It is not excluded that a further subdivision of the regions (using clusters for example) is operated. In any case, our programs will be updated in such a case.\\

\noindent (e) In each household, the member are numbered from one to the total size of the household. We do not give a name to this variable. It will we denoted by 
$i$ in the packages.\\

\noindent The three strata variables \textit{Region}, \textit{District} [and cluster if present] and \textit{Household} will be combined to aggregated the data at the household level.\\

\bigskip \noindent \textbf{C - Variables}.\\

Each \textbf{HDB} includes a more or less large number of variables, which themselves depend of the objectives of the survey.  For example, the variables in the \textbf{HDB ESAM 1} or \textbf{ESAM 2} of Senegal are quite different from those the \textit{HDB ELIM 1} of Mali.\\

\noindent In future papers, we will be dealing with specific variables.\\

\section{Income or Expense data}

In this section, we learn how to create aggregated data at the level of the household from the person data file. Next, we will practice on real databases and provide successfully tested computer programs  

\bigskip \noindent \textbf{A - Aggregation and equivalence scales}.\\

\noindent Aggregating the data at the household level is a very important matter. In this paper, we will focus in quantitative analysis. We will have to deal with equivalence scales.\\

\noindent But before we proceed to it, we want to mention that aggregating categorical variables in the available data offers many opportunities to conduct interesting research based on random measures and on symbolic data analysis. For example, the age variable may not only be aggregated into a median or a mean. Instead, it would be interesting to considered it as a distribution, for example,  on $D=[0,100]$, and then to study random distributions. This kind of approach would open fruitful researches in our Statistics and Mathematics departments in relation of the development of very new trends of the theory.\\

\noindent Let us go back to the quantitative aggregation problems, especially concerning income or expense data.\\

We will not enter into the theoretical details on the adult-equivalence question. We will remain at a simple description level of the question. We want to be able to compare different groups on persons in terms of life quality, when the latter is measured by the income. If we decide to use the income variable as the measure of life quality, the problem is how to compare two groups of people, namely two households, in terms of the income of the members of each group despite of the difference of their structure, the difference of the contexts and perhaps, of the different environments.\\

\noindent To simplify, if we compare two households, each of them reduced to one adult person, certainly their incomes may help. Suppose now that we have two households such that :\\

\noindent (1) the first includes two male adults and a female adult,\\

\noindent (2) and the second is formed by a female single parent and one 10 year old girl. It is not fair to compare the well-being of these households by the total income of each of them.\\

\noindent One has to find a way to convert the individual incomes  of the different members of one household to the income of one adult male person. This income is called the adult-equivalent income of the household. Finding the adult-equivalent incomes for all households using the same method would, actually, allow a fair comparison of the household.\\

\noindent A number of methods for providing such adult-equivalent incomes, in the form of scales, are available. We may cite some authors who gave interesting contributions in that topic : \cite{pollack}, \cite{bradbury}, \cite{duclos}, \cite{hourriez} and the references therein. Let $h$ denote some household of size $H$. An adult-equivalence scale may be defined as an allocation of some weight $w(i,h)$, which is a number between zero and one, to each member of the household $h$ and the income $I(i,h)$ of the $i-th$ member contribute the amounts $w(i,h)I(i,h)$ to the global adult-equivalent income of the households $h$, which is

$$
X(h)=w(1,h)I(1,h)+\cdots+w(H,h)I(H,h).
$$  

\noindent The weight $w(i,h)$ is justified in each proposed scale in very elaborated researches (see already cited papers). The common scales mainly use the gender and the age. Usually, people older than 15 years are considered as adults. If not they are children. We present some common scales.\\

\noindent (a) \textbf{Oxford Scale}. This scale allocates the full weight to adult household chief and the weight $0.7$ to other adults, and the weight $0.5$ to non-adults.\\

\noindent (b) \textbf{FAO and OMS scale}. Here, the following allocation is used. Male adults : 1. Female adults : 0.8. Non-adults : 0.5.\\

\noindent (b) \textbf{Duclos and Mercadier-Pats (DMP)} scale. Denote by $N_{a}(h)$ the number of adults and $N_{e}(h)$ the number of non-adults in the household $h$ and consider two real numbers $c$ and $s$ in $[0,1]$.  For any choice of $c$ and $s$, form

$$
E(h)=(N_{a}(h)+cN_{e}(h))^{s}
$$

\noindent The adult-equivalent income of the household $h$ is the total income of the household divided by $E(h)$. We notice here that the \textit{DMP} scale cannot be applied directly to the individual, to the contrary of the \textit{Oxford} and \textit{Fao-Fam} scales.\\

\section{Practice}

\noindent This version will focus on \textit{MS VB}\texttrademark \ codes. We recommend to start by Open the \textit{MS Excel}\texttrademark \ or the \textit{IBM SPSS}\texttrademark, and for each variable we want to use :\\

\noindent (1) Copy the associated column\\
\noindent (1) Open a text file (with Notepad, Bloc note, etc.)\\
\noindent (1) Paste the copied column\\
\noindent (1) Name the file and save it.\\

\noindent Using this method, we create the files : \textbf{region.text} (for the main area division), \textbf{milieu.txt}, \textbf{cluster.txt}, \textbf{household.txt}, \textbf{gender.txt}, \textbf{age.txt} holding the variables of the same names : region, milieu, cluster, household, gender, age.\\

\noindent We also need the variable \textbf{poswrchief} (position with respect to the chief of the household), which indicates the relation of the member with respect to the household chief. Usually, the label one is allocated to the household chief. We store this variable in the file \textbf{poswrchief.txt}

\noindent In the sequel, we will comment the algorithms and computer packages provided in the Packages Section \ref{packages}.\\

\bigskip \noindent \textbf{Mali's ELIM Database}.\\

\noindent \textbf{Step 1}. Identification of the household.\\

\noindent In this paper, we explain the methods using the \textbf{Mali ELIM 1 database}. The programs we successfully used will be given here in Section 
\ref{packages}.
All the scripts are gathered in a file you can find in \cite{extern}. However, this external resources file will be updated to include other applications.\\

\noindent In some \textit{HDB}'s, the households are numbered for the one to the last one. In some others, the number is re-initialized in each district or cluster. In all cases, the following method works.\\

\noindent In the presence case, it is enough to use an identification in the form :

\noindent R+region+M+milieu+C+cluster+H+household.\\

\noindent We use the procedure \textit{householdIdentifier()} provided in \textbf{(01) - Identification} in Section \ref{packages} to get the variable \textbf{identhousehold} restored in the file \textbf{identhousehold.txt}.\\

\noindent Since, this is our first use of a procedure or a function, we want to draw the attention of the readers on the following facts.\\

\noindent (a) A procedure or a function is written in a way that it allows the reader to render the program into any software he knows better. Lines beginning by \textbf{Rem} are commentaries, in which we explain what do the codes below. So, a reader will be able to adapt these programs into a language he is acquainted with.\\

\noindent (b) In all the packages presented here, the files are supposed to be in a folder that is specified in Line 7 of the procedure in \textbf{(01) - Identification)}. The ready has to pay attention to this, to make sure his programs run correctly.\\

\bigskip \noindent \textbf{Step 2}. Construction of adult-equivalence scales Functions.\\

\noindent Here, we should be aware of the different presentations of the age variables. Like in Mali, the age of an individual is the real age, for some months (a fraction of year) to maximum age in years. \textbf{In ELIM 1}, the code 99 is reserved to an unknown age. This is a clear indication that the survey did not meet an older person that 99 years in Mali during the survey. \\

\noindent In some databases, like in \textbf{ESAM 1 of Senegal}, the age is given by classes of five years. For example, the first class $]0,5[$ may be denoted by $1$,  $[5,10[$ by 2, $[10,14[$ by $3$, $[5,10[$ by $5$, etc.\\

\noindent As well, the labels of the gender, male next female, may be $(0,1)$ or $(1,2)$.\\

\noindent Rather than constructing directly the equivalence-adult from the \textit{person data file}, we propose functions scale for the \textbf{FAO-FAM} and the \textbf{Oxford} scales to be used as subroutines. We notice that this is not possible for the DMP scale.\\
  
\noindent The functions given here, use a discriminatory variable \textbf{TypeAge}. If the age is given in years, we replace \textbf{TypeAge} by $1$. We replace it by $2$ if the age is given by classes of 5 year range. If the age is not provided in any of these forms, the functions are set to 0.99, which is an error code.\\ 

\noindent In the same manner, the variable \textbf{TypeGender} is used. If the gender is organized on the basis $(0=Male, 1=Female)$, we replace \textbf{TypeGender} by $1$. We replace it by $2$ if the gender is codifies as $(1=Male, 2=Female)$. If the gender is not provided in any of these forms, the functions are set to 0.99, which is an error code.\\ 

\noindent The adult-equivalence scales functions for the \textit{Oxford} scales and the \textit{Fao-Fam} are given Subsection \textbf{02 - Scales Oxford, Fao-Fam Functions} in the packages section \ref{packages}.\\

\noindent We cannot have a \textit{DPM scale} function at the level of individuals, since we have to compute the total number of children and the number of adults in a household. This scale is based on the household while the \textit{Oxford} and the \textit{Fao-Fam} scales are based on individuals.\\

\noindent We will create the file of \textit{DPM} scaling at the level of the households.\\

\bigskip \noindent \textbf{Step 3}. Construction of adult-equivalence scales files.\\

\noindent (A) To create the \textit{Oxford} scaling file, we proceed from the persons files as follows :\\

\noindent (1) Open the identification files for households (\textbf{identhousehold.txt}, created at Step 1) [A], the age file [B], gender file [C]: for reading. Open the Oxford scaling file [D], and the household size file [E], for exporting. Initialize \textbf{NBRH=1}. Initialize the household size $\textbf{HSIZE}=1.$\\ 

\noindent (2) Read the first line of the files [A], [B] and [C]. Compute the Oxford scale for this individual and initialize the scale value \textbf{OxfordScaleCurrent} for this household \textbf{HCurrent}.\\

\noindent (3) Continue reading the following lines. As long as the households is the same, increment the scale value \textbf{OxfordScaleCurrent} for this household \textbf{HCurrent} with the scale value computed at the corresponding line. And increment \textbf{HSIZE} by one.\\

\noindent (4) At the first line where the household is no longer \textbf{HCurrent} :\\

\noindent \quad (4.1) The current scale value \textbf{OxfordScaleCurrent} is the Oxford scale for that household \textbf{HCurrent}. Export the value to the file [D]. Export \textbf{HSIZE} to [E].\\

\noindent \quad (4.2) Give to \textbf{HCurrent} the value of household that is read at that line, and initialize its \textbf{OxfordScaleCurrent} at the scale value that is computed at that line. Initialize \textbf{HSIZE} to one. \\

\noindent \quad (4.3) Increment \textbf{NBRH}.\\

\noindent (5) Repeat steps (3) and (4).\\

\noindent (6) At the end the procedure, \textbf{HCurrent} is the last household, \textbf{NBRH} is the number of households. \textbf{OxfordScaleCurrent} is the scale value of the last household. Export it to [D] and \textbf{HSIZE} to [E].\\

\noindent (7) Close all the files.\\

\noindent At the end, we get the \textit{Oxford scale file} for the database. (See the subroutine \textbf{createFilesScaleOxford()} in Sub-subsection 
\textbf{03 (a) Oxford Scale File} in in Section \ref{packages}, for the package) \\

\bigskip \noindent (B) To create the \textit{Fao-Fam} scaling file, we proceed similarly. But we have to add the file on the status of the individual in the household (\textbf{powsrchief}, which is one for the household chief and greater that one for the others).\\

\noindent And we get the \textit{fao-fam scale file}. (See the subroutine \textbf{createFilesFaoFam()} in Sub-subsection\textbf{03 (b) Oxford Fao-Fam File} in Section \ref{packages}, for the package)\\

\noindent (C) to create the \textit{DMC} scaling file, we use the following algorithm.\\

\noindent (1) Fix the values $c$ and $s$ (in the example : $s=0.5$, $c=0.7$).\\

\noindent (2) Open the identification files for households (\textbf{identhousehold.txt}, created at Step 1) [A], the age file [B] : for reading. Open the Oxford scaling file [C], for exporting. Initialize \textbf{NBRH=1}.\\ 

\noindent (3) Read the first line of the files [A] and [B]. Initialize number of children $NBRC=0$ and the number of adults $NBRA=0$. If the current individual is a children, increment $NBRC=0$ by one, otherwise increment $NBRA=0$ by one. Name the current household by \textbf{HCurrent}.\\

\noindent (4) Continue reading the following lines. As long as the households is the same, increment $NBRC=0$ by one if the current individual is a children, otherwise increment $NBRA=0$ by one.\\

(5). At the first line where the household is no longer \textbf{HCurrent} :\\

\noindent \quad (5.1) The current values of $NBRC=0$ and $NBRA=0$  are respectively the number of Children and the number of adults. Export the value
$$
(NBRA+c \times NBRA)^{s}
$$

\noindent \noindent to [C].\\

\noindent \quad (5.2) Give to \textbf{HCurrent} the value of household that is read at that line.\\

\noindent \quad (5.3) Initialize number of children $NBRC=0$ and the number of adults $NBRA=0$. If the current individual is a children, increment $NBRC=0$ by one, otherwise increment $NBRA=0$ by one.\\

\noindent (6) Repeat steps (4) and (5).\\

\noindent (7) At the end the procedure, \textbf{HCurrent} is the last household, \textbf{NBRH} is the number of households. Export the value of Point (4.1) to [C].\\

\noindent (8) Close all the files.\\

\noindent At the end, we get the \textbf{DMP} scale file file for the database. (See the subroutine \textbf{createFilesDMP()} in
 Sub-subsection \textbf{03 (c) DMC Scale File} in Section \ref{packages}, for the package)\\

\bigskip \noindent \textbf{Step 4}. Construction of total income or expense files.\\

\noindent First, let us consider the case where the income of any member of the file is available (monthly or yearly). We will use the same algorithm that was used in Step 3. We proceed as follows.\\

\noindent (1) Open the identification files for households (\textbf{identhousehold.txt}, created at Step 1) [A], the income file [B]: for reading. Open the \textbf{totalincome} file [C], for exporting. \\ 

\noindent (2) Read the first line of the files [A] and [B]. Initialize the total income of the household to \textbf{TotalIncomeCurrentHousehold} to the value read in the file [B].\\

\noindent (3) Continue reading the following lines. As long as the households is the same, increment \textbf{TotalIncomeCurrentHousehold} by the value read from the file [B].\\

\noindent (4) At the first line where the household is no longer \textbf{HCurrent} :\\

\noindent \quad (4.1) The current value of \textbf{TotalIncomeCurrentHousehold} is the total income for that household \textbf{HCurrent}. Export the value in the file [C].\\

\noindent \quad (4.2) Give to \textbf{HCurrent} the value of household that is read at that line, and initialize \textbf{TotalIncomeCurrentHousehold} to the value read in the file [B].\\

\noindent (5) Repeat steps (3) and (4).\\

\noindent (6) At the end the procedure, \textbf{HCurrent} is the last household. Here, \textbf{TotalIncomeCurrentHousehold} is the total income the last household. Export it to [C].\\

\noindent (7) Close all the files.\\

\noindent In implementing this algorithm, we may have different variants. Let us highlight them with real examples.\\

\bigskip \noindent In the \textit{ELIM1} \textit{HDB} of Mali, the only variable giving information on the income amount is the amount earned in the month before the survey. And the variable is an categorical one with values $A$, $B$, ..., $L$ corresponding to income ranges $0 - 29.000$, $29.001 - 100.000$, ...., $\geq 3.500.000$. In such a situation we have to transform this variable into a new numerical new. In the study, we created a new variable for which every individual is assigned the mean value of the range income into which he falls. For the last range, we assign the fixed income $3.000.000$ to corresponding individuals. The associated program is below at \textbf{ (6) Example of categorical income variable : case of ELIM1, Mali}.\\

\noindent If the case of \textit{ESAM 2} of Senegal, the income is not available, but the expense is present and broken into different categories such such as : health, education expenses. In this case, we should add to the algorithm all the expenses variable and use their sums at the place of the one value of the income.\\

\noindent \textbf{Application to the case \textit{ELIM1} of Mali}. Based on the transformed income variable exported in the file \textbf{totalincome.txt}, the subroutine
in Subsection \textbf{(4) Total income File} in Section \ref{packages}.\\

\bigskip \noindent \textbf{Conclusion}. Now, we have workable files from which we may do statistics works in the aggregated data. In the example of Mali's \textit{ELIM1} database, we have :\\

\noindent (1) The total income data of the households.\\

\noindent (2) The adult-equivalence files (Oxford and Fao-Fam) for scaling the income. One example of a \textit{DMP} adult-equivalence file is provided for particular cases of $c=0.5$ and $s=0.7$. But the package is there to do the same for values of $c$ and $s$.\\

\noindent (3) The scale income $income/scale$ be be used for statistical studies.\\

\bigskip \noindent \textbf{Step 4}. One we have the scaled income for the households, we might be interested by applying of it many procedures using label data. This is interesting since most of the variables in \textit{HDB}'s are categorical : the area, the level education, the profession, etc. To create any label variable for households based some variable, we proceed as previously.\\

\noindent Let us give two examples.\\

\noindent \textbf{Algorithm for the Area label file}.\\

\noindent (1) Open the identification files for households (\textbf{identhousehold.txt}, created at Step 1) [A], the gender file [B]: for reading. Open the \textbf{label area} file [C], for exporting. \\ 

\noindent (2) Read the first line of the files [A] and [B]. Take the read area value as area label of the current household. Export it to [C].\\

\noindent (3) Continue reading the following lines as long as the households is the same.\\

\noindent (4) At the first line where the household is no longer \textbf{HCurrent} :\\

\noindent \quad (4.1) Take the area value at that line as the area label of the current current household. Export it to [C].\\

\noindent \quad (4.2) Give to \textbf{HCurrent} the value of household that is read at that line.\\

\noindent (5) Repeat steps (3) and (4).\\

\noindent (6) Close all the files.\\

\bigskip The corresponding subroutine is \textbf{createLabelVariableHouseHoldB()}. It is given in Sub-subsection \textbf{(5a) Creation of the area label file} in Section \ref{packages}.\\

\noindent \textbf{Algorithm for the household chief area label file}.\\

\noindent This case is more elaborated since we have we have to combine with the individual status in the household (powsrchief, which is one for the household chief and greater that one for the others).\\

\noindent (1) Open the identification files for households (\textbf{identhousehold.txt}, created at Step 1) [A], the area (region) file [B], the \textbf{poswrchief} file [C] : for reading. Open the \textbf{Chief gender label} file [D], for exporting. \\ 

\noindent (2) Read the first line of the files [A] and [B]. Initialize \textbf{labelchiefgender} to XXX. If the value read from the \textbf{poswrchief} file is "1", take the read gender as the gender label of the current household and name it \textbf{labelchiefgender}.\\

\noindent (3) Continue reading the following lines. As long as the households is the same : \\
\quad If the value read from the \textbf{poswrchief} file is "1", take the read gender as the gender label of the current household and name it \textbf{labelchief gender}. If not, do nothing.\\

\noindent (4) At the first line where the household is no longer \textbf{HCurrent} :\\ 

\noindent \quad (4.1) Export \textbf{labelchiefgender} to file [D].\\

\noindent \quad (4.2) Give to \textbf{HCurrent} the value of household that is read at that line.
\noindent \quad (4.3) Initialize \textbf{labelchiefgender} to "XXX". If the value read from the \textbf{poswrchief} file is "1", take the read gender as the gender label of the current household and name it \textbf{labelchief gender}.\\

\noindent (5) Repeat steps (3) and (4).\\

\noindent (6) If \textbf{labelchiefgender} is not "XXX", then export it to file [D].\\

\noindent (7) Close all the files.\\

\bigskip \noindent The corresponding subroutine is \textbf{createLabelVariableHouseHoldA()}. It is given in Sub-subsection \textbf{(5b) Creation of the Household Chief Gender label file} in Section \ref{packages}.\\

\bigskip \noindent \textbf{}The same packages given in \textbf{MS VB\texttrademark} \ are written in \textbf{R} codes. The only difference is that the files are not created. In the script file, the variables are created and are ready to be used in statistical procedures. The reader may find it as well in \cite{extern}.\\

\section{Conclusions} The papers showed how to proceed to handle households databases to extract workable quantitative random variables and label variables on which different statistical procedures may be applied. In the particular case where we are interested by the total income file, different adult-equivalence files and label variables such as the geographical label, the chief gender label have been created. The packages in Visual Basic\texttrademark \ and in the free \textit{R} software are provided to serve as models. The descriptions of the algorithms should allow to adapt them to any household database and to arbitrary databases. Their implementation in \textbf{Oracle Java}\texttrademark \ , \textbf{Python}, \textbf{PHP}  etc., are easily set up. In the future, we will be able to deliver reports on some more or less complex studies using the available databases. It will be meant that the packages we will use are based on the current ones.

\newpage 
\section{Packages} \label{packages}

\noindent \textbf{(01) - Identification}.

\begin{lstlisting}
Sub householdIdentifier()

Dim region As String, household As String, cluster As String, 
Dim milieu As String, folder As String
Dim chain As String, newchain As String

Rem Use the right folder where the file are located
folder = "mali/elim1/"

Rem files in text formats, for getting the data
region = "region.txt"
household = "menage.txt"
cluster = "cluster.txt"
milieu = "milieu.txt"

Rem new file to get the identification
identhousehold = "identhousehold.txt"

Rem open the files : region, household, cluster, milieu, for reading
Open folder \& region For Input As \#1
Open folder \& milieu For Input As \#2
Open folder \& cluster For Input As \#3
Open folder \& household For Input As \#4


Rem open the files : identhousehold.txt, for export
Open folder \& identhousehold For Output As \#5

While Not EOF(1)
    newchain = ""
    Input \#1, chain:   newchain = newchain \& "D" \& chain
    Input \#2, chain:   newchain = newchain \& "M" \& chain
    Input \#3, chain:   newchain = newchain \& "C" \& chain
    Input \#4, chain:   newchain = newchain \& "H" \& chain
    
    Print \#5, newchain
Wend
Close \#1, \#2, \#3, \#4, \#5
MsgBox "identhousehold.text ready to be used in your folder!"
End Sub
\end{lstlisting}

\newpage
\noindent \textbf{02 - Scales Oxford, Fao-Fam Functions}.\\

\noindent \textbf{02 (a) Oxford Scale}.\\

\begin{lstlisting}

Function scaleOxford(ageType As Integer, age As Double, poswtchief As Integer) As Double

Rem ageType (1 for a real age in year), (2 if the age is given in classes of length of five years)
Rem The value poswtchief is one for the household chief. The alternate value is not relevant
Rem If one the arguments is out of range, the scale is set to 0.99, as an error code

If (ageType < 1 Or ageType > 2) Then
    scaleOxford = 0.99
Else
    Select Case ageType
        Case 1:
            If (age < 15) Then
                    scaleOxford = 0.5
            Else
                    If (poswtchief = 1) Then
                    scaleOxford = 1
                    Else
                    scaleOxford = 0.7
                    End If
            End If
            
        Case 2:
            
            If (age < 4) Then
                    scaleOxford = 0.5
            Else
                    If (poswtchief = 1) Then
                    scaleOxford = 1
                    Else
                    scaleOxford = 0.7
                    End If
            End If

End Select
End If

End Function
\end{lstlisting}

\newpage \noindent \textbf{02 (b) Fao-FAM Scale}.\\
\begin{lstlisting}
Function scaleFaoFam(ageType As Integer, genderType As Integer, age As Double, gender As Integer)
Rem ageType (1 for a real age in year), (2 if the age is given in classes of length of five years)
Rem genderType (1 for Male=0, female=1), (2 for Male=1, female=2)
Rem If one the argument is out of range, the scale is set to 0.99, as an error code

Dim testContinuation As Boolean

testContinuation = False

If ((ageType < 1) Or (ageType > 2) Or (genderType < 1) Or (genderType > 2)) Then
    scaleFaoFam = 0.99
Else
    Select Case ageType
        Case 1:
            If (age < 15) Then
                scaleFaoFam = 0.5
            Else
                testContinuation = True
                
            End If
           
        Case 2:
            If (age < 4) Then
                scaleFaoFam = 0.5
            Else
                testContinuation = True
            End If

End Select

End If

If (testContinuation) Then
            Select Case genderType
                Case 1:
                    If (gender = 0) Then
                        scaleFaoFam = 1
                    Else
                            If (gender = 1) Then
                                scaleFaoFam = 0.8
                            Else
                                scaleFaoFam = 0.99
                            End If
                    End If
                    
                Case 2:
                
                    If (gender = 1) Then
                        scaleFaoFam = 1
                    Else
                            If (gender = 2) Then
                                scaleFaoFam = 0.8
                            Else
                                scaleFaoFam = 0.99
                            End If
                    End If
            End Select
End If


End Function
\end{lstlisting}

\newpage
\noindent \textbf{03 - Creating Scales Files}.\\

\noindent \textbf{03 (a) Oxford Scale File}.\\
\begin{lstlisting}

Sub createFilesScaleOxford()

Dim identhouseholdfile As String, agefile As String, genderfile As String, poswrchieffile As String, sizehouseholdfile As String, scaleoxfordfile As String
Dim i As Double, j As Double, household As String, age As Double, gender As Integer, poswrchief As Integer, agesrt As String, gendersrt As String, poswrchiefsrt As String
Dim ageType As Integer, genderType As Integer, scaleValue As Double, nbrhouseholds As Double, sizehousehold As Double
Dim folder As String

Rem provide Type of Age / Type of Gender
Rem ageType (1 for a real age in year), (2 if the age is given in classes of length of five years)
Rem genderType (1 for Male=0, female=1), (2 for Male=1, female=2)
Rem If one the argument is out of range, the scale is set to 0.99, as an error code
ageType = "1"
genderType = "1"

Rem Test if Type of Age / Type of Gender are correct
If ((ageType < 1) Or (ageType > 2) Or (genderType < 1) Or (genderType > 2)) Then
    MsgBox "Type of Age and/or Type of Gender wrong Fix it Begin again"
    Exit Sub
End If


Rem Use the right folder where the file are located
folder = "C:/Data_gslo/gslo/bdm/bdm/packagevb6/mali/elim1/"


Rem files needed in text formats : identhousehold , age, gender, poswrchief, sizehousehold, scaleOxford
identhouseholdfile = "identhousehold.txt"
agefile = "age.txt"
genderfile = "gender.txt"
poswrchieffile = "poswrchief.txt"

Rem new file to get the identification
scaleoxfordfile = "scaleoxford.txt"
sizehouseholdfile = "sizehousehold.txt"

Rem open the files : region, household, cluster, milieu, for reading
Open folder & identhouseholdfile For Input As #1
Open folder & agefile For Input As #2
Open folder & poswrchieffile For Input As #3
Open folder & genderfile For Input As #6


'open the files : identhousehold, for export
Open folder & scaleoxfordfile For Output As #4
Open folder & sizehouseholdfile For Output As #5

'rem read the first line and initialize scaleValue
    Input #1, household
    Input #2, agestr
    Input #3, poswrchiefstr
    Input #6, genderstr
    
    
    poswrchief = Int(Val(poswrchiefstr))
    age = Val(agestr)
    
    Rem Catch the current household
    householdCurrent = household
    
    Rem initialize scale value for the current household
    scaleValue = scaleOxford(ageType, age, poswrchief)
    
    
    Rem Initialize the size of the household at j=1
    sizehousehold = 1
    Rem Number of households
    nbrhouseholds = 1
    
    j = 1
    
    
     While Not EOF(1)
     j = j + 1
     Input #1, household
     Input #2, agestr
     Input #3, poswrchiefstr
     Input #6, genderstr
     poswrchief = Int(Val(poswrchiefstr))
     age = Val(agestr)
     
     
     If (household = householdCurrent) Then
    
           Rem As long as the household is the same, we contuinue to increment the scale and the size
           scaleValue = scaleValue + scaleOxford(ageType, age, poswrchief)
           sizehousehold = sizehousehold + 1
    Else
    
        Rem The first time we did quit the current household, we may save the current value
        Rem of scaleValue as the scale of the the the household which is just completed
        Print #4, Trim(Str(scaleValue))
        Print #5, sizehousehold
        
        Rem Initialize the sclale and the size for the new household
        scaleValue = scaleOxford(ageType, age, poswrchief)
        Rem Initialize the size of the household at j=1
        sizehousehold = 1
        Rem inceremente number of househols
        nbrhouseholds = nbrhouseholds + 1
        householdCurrent = household
    End If
    
Wend
        Print #4, Trim(Str(scaleValue))
        Print #5, sizehousehold

Close #1, #2, #3, #4, #5, #6
MsgBox newchain & " number of households = " & nbrhouseholds & " Files scaleoxford.txt and sizehousehold.txt ready to be used"
End Sub
\end{lstlisting}

\noindent \textbf{03 (b) Fao-Fam Scale File}.\\

\begin{lstlisting}
Sub createFilesScaleFaoFam()

Dim identhouseholdfile As String, agefile As String, genderfile As String, poswrchieffile As String, sizehouseholdfile As String, scaleoxfordfile As String
Dim i As Double, j As Double, household As String, age As Double, gender As Integer, poswrchief As Integer, agesrt As String, gendersrt As String, poswrchiefsrt As String
Dim ageType As Integer, genderType As Integer, scaleValue As Double, nbrhouseholds As Double, sizehousehold As Double
Dim folder As String

Rem provide Type of Age / Type of Gender
Rem ageType (1 for a real age in year), (2 if the age is given in classes of length of five years)
Rem genderType (1 for Male=0, female=1), (2 for Male=1, female=2)
Rem If one the argument is out of range, the scale is set to 0.99, as an error code
ageType = "1"
genderType = "2"

Rem Test if Type of Age / Type of Gender are correct
If ((ageType < 1) Or (ageType > 2) Or (genderType < 1) Or (genderType > 2)) Then
    MsgBox "Type of Age and/or Type of Gender wrong Fix it Begin again"
    Exit Sub
End If


Rem Use the right folder where the file are located
folder = "C:/Data_gslo/gslo/bdm/bdm/packagevb6/mali/elim1/"


Rem files needed in text formats : identhousehold , age, gender, poswrchief, sizehousehold, scaleOxford
identhouseholdfile = "identhousehold.txt"
agefile = "age.txt"
genderfile = "gender.txt"
poswrchieffile = "poswrchief.txt"

Rem new file to get the identification
scalefaofamfile = "scalefaofam.txt"
sizehouseholdfile = "sizehousehold.txt"

Rem open the files : region, household, cluster, milieu, for reading
Open folder & identhouseholdfile For Input As #1
Open folder & agefile For Input As #2
Open folder & poswrchieffile For Input As #3
Open folder & genderfile For Input As #6


'open the files : identhousehold, for export
Open folder & scalefaofamfile For Output As #4
Open folder & sizehouseholdfile For Output As #5

'rem read the first line and initialize scaleValue
    Input #1, household
    Input #2, agestr
    Input #3, poswrchiefstr
    Input #6, genderstr
    
    
    poswrchief = Int(Val(poswrchiefstr))
    gender = Int(Val(genderstr))
    age = Val(agestr)
    
    Rem Catch the current household
    householdCurrent = household
    
    Rem initialize scale value for the current household
    scaleValue = scaleFaoFam(ageType, gender, age, poswrchief)
    
    Rem Initialize the size of the household at j=1
    sizehousehold = 1
    Rem Number of households
    nbrhouseholds = 1
    
    j = 1
    
    
     While Not EOF(1)
     j = j + 1
     Input #1, household
     Input #2, agestr
     Input #3, poswrchiefstr
     Input #6, genderstr
     
     Rem Transform strings into integers or double
     poswrchief = Int(Val(poswrchiefstr))
     age = Val(agestr)
     gender = Int(Val(genderstr))
     
     If (household = householdCurrent) Then
    
           Rem As long as the household is the same, we contuinue to increment the scale and the size
           scaleValue = scaleValue + scaleFaoFam(ageType, gender, age, poswrchief)
           sizehousehold = sizehousehold + 1
    Else
    
        Rem The first time we did quit the current household, we may save the current value
        Rem of scaleValue as the scale of the the the household which is just completed
        Print #4, Trim(Str(scaleValue))
        Print #5, sizehousehold
        
        Rem Initialize the sclale and the size for the new household
        scaleValue = scaleFaoFam(ageType, gender, age, poswrchief)
        Rem Initialize the size of the household at j=1
        sizehousehold = 1
        Rem inceremente number of househols
        nbrhouseholds = nbrhouseholds + 1
        householdCurrent = household
    End If
    
Wend

Print #4, Trim(Str(scaleValue))
Print #5, sizehousehold
        
Close #1, #2, #3, #4, #5, #6
MsgBox j & "/ Number of households = " & nbrhouseholds & " Files scalefaofam.txt and sizehousehold.txt ready to be used"
End Sub

\end{lstlisting}

\noindent \textbf{03 (c) DMP Scale File}.\\
\begin{lstlisting}

Sub createFilesScaleDMP()

Dim identhouseholdfile As String, agefile As String, genderfile As String, poswrchieffile As String, sizehouseholdfile As String, scaleoxfordfile As String
Dim i As Double, j As Double, household As String, age As Double, gender As Integer, poswrchief As Integer, agesrt As String, gendersrt As String, poswrchiefsrt As String
Dim ageType As Integer, genderType As Integer, scaleValue As Double, nbrhouseholds As Double, sizehousehold As Double
Dim folder As String, scaleChildren As Double, scaleAdult As Double, CDPM, SDPM As Double

Rem provide Type of Age / Type of Gender
Rem ageType (1 for a real age in year), (2 if the age is given in classes of length of five years)
Rem genderType (1 for Male=0, female=1), (2 for Male=1, female=2)
Rem If one the argument is out of range, the scale is set to 0.99, as an error code
ageType = "1"


Rem Give the values if c and s for the Duclos-Pats-Mercadier
CDPM = 0.5
SDPM = 0.7

Rem Test if Type of Age / Type of Gender are correct
If ((ageType < 1) Or (ageType > 2)) Then
    MsgBox "Type of Age wrong! Fix it Begin again"
    Exit Sub
End If


Rem Use the right folder where the file are located
folder = "C:/Data_gslo/gslo/bdm/bdm/packagevb6/mali/elim1/"


Rem files needed in text formats : identhousehold , age, gender, poswrchief, sizehousehold, scaleOxford
identhouseholdfile = "identhousehold.txt"
agefile = "age.txt"

Rem new file to get the identification
scaleDPMfile = "scaleDPM-" & Trim(Str(CDPM)) & "-" & Trim(Str(SDPM)) & ".txt"
sizehouseholdfile = "sizehousehold.txt"

Rem open the files : region, household, cluster, milieu, for reading
Open folder & identhouseholdfile For Input As #1
Open folder & agefile For Input As #2

MsgBox "lo" & scaleDPMfile
'open the files : identhousehold, for export
Open folder & scaleDPMfile For Output As #4

'rem read the first line and initialize scaleValue
    Input #1, household
    Input #2, agestr
    
    
    age = Val(agestr)
    
    Rem Catch the current household
    householdCurrent = household
    
    Rem initialize scale value for the current household
    scaleChildren = 0: scaleAdult = 0
    If (ageType = 1) Then
                If (age < 15) Then scaleChildren = 1 Else scaleAdult = 1
    Else
                If (age < 4) Then scaleChildren = 1 Else scaleAdult = 1
    End If
    
    Rem Initialize the size of the household at j=1
    sizehousehold = 1
    Rem Number of households
    nbrhouseholds = 1
    
    j = 1
    
    
     While Not EOF(1)
     j = j + 1
     Input #1, household
     Input #2, agestr
     
     
     Rem Transform strings into integers or double
     poswrchief = Int(Val(poswrchiefstr))
     age = Val(agestr)
     gender = Int(Val(genderstr))
     
     If (household = householdCurrent) Then
    
           Rem As long as the household is the same, we contuinue to increment the scale and the size
           
            
            If (ageType = 1) Then
                If (age < 15) Then scaleChildren = scaleChildren + 1 Else scaleAdult = scaleAdult + 1
            Else
                If (age < 4) Then scaleChildren = scaleChildren + 1 Else scaleAdult = scaleAdult + 1
            End If
           
            sizehousehold = sizehousehold + 1
    Else
    
        Rem The first time we did quit the current household, we may save the current value
        Rem of scaleValue as the scale of the the the household which is just completed
        Print #4, Trim(Str((scaleAdult + (CDPM * scaleChildren)) ^ (SDPM)))
        
        
        Rem Initialize the sclale and the size for the new household
        scaleChildren = 0: scaleAdult = 0
        If (ageType = 1) Then
                If (age < 15) Then scaleChildren = 1 Else scaleAdult = 1
        Else
                If (age < 4) Then scaleChildren = 1 Else scaleAdult = 1
        End If
        
        Rem Initialize the size of the household at j=1
        sizehousehold = 1
        Rem inceremente number of househols
        nbrhouseholds = nbrhouseholds + 1
        householdCurrent = household
    End If
    
Wend
Print #4, Trim(Str((scaleAdult + (CDPM * scaleChildren)) ^ (SDPM)))
Close #1, #2, #4
MsgBox "Number of households = " & nbrhouseholds & " Files scalefaofam.txt and sizehousehold.txt ready to be used"
End Sub

\end{lstlisting}

\newpage
\textit{(4) Total monthly income file}
\begin{lstlisting}

Sub createTotalIncomeMali()

Dim identhouseholdfile As String, household As String
Dim incomestr As String, income As Double, incomefile As String, incometotalfile As String, incomeHousehold As Double

Rem Use the right folder where the file are located
folder = "C:/Data_gslo/gslo/bdm/bdm/packagevb6/mali/elim1/"


Rem files needed in text formats : identhousehold , monthlyincome
identhouseholdfile = "identhousehold.txt"
incomefile = "monthlyincome.txt"


'open the files : to receive Total Income/Household
incometotalfile = "totalincome.txt"

identhouseholdfile = "identhousehold.txt"
Open folder & identhouseholdfile For Input As #1
Open folder & incomefile For Input As #2
Open folder & incometotalfile For Output As #3

'rem read the first line and initialize scaleValue
    Input #1, household
    Input #2, incomestr
    
    income = Val(incomestr)
    
    Rem Catch the current household
    householdCurrent = household
    
    Rem initialize the value of the income
    incomeHousehold = income
    
    
    j = 1
    
    
     While Not EOF(1)
     j = j + 1
     Input #1, household
     Input #2, incomestr
     income = Val(incomestr)
     
     
     
     If (household = householdCurrent) Then
    
            Rem As long as the household is the same, we contuinue to increment the scale and the size
           
            incomeHousehold = incomeHousehold + income
    
    Else
    
        Rem The first time we did quit the current household, we may save the current value
        Rem of scaleValue as the scale of the the the household which is just completed
        Print #3, incomeHousehold
        
        
        Rem Initialize the sclale and the size for the new household
                
                householdCurrent = household
                incomeHousehold = income
    End If
    
Wend

Close #1, #2, #3
MsgBox "Total Incom found and created in incometotal.txt, and ready to be used"
End Sub

\end{lstlisting}

\newpage
\noindent \textbf{(5) Example of creation of a label variable}.\\

\noindent \textbf{(5a) Creation of the area label file}.\\

\begin{lstlisting}
Sub createLabelVariableHouseHoldB()

Dim identhouseholdfile As String, household As String
Dim regionfile As String, labelregionfile As String, region As String


Rem Use the right folder where the file are located
folder = "C:/Data_gslo/gslo/bdm/bdm/packagevb6/mali/elim1/"


Rem files needed in text formats : identhousehold , gender, poswrchief
identhouseholdfile = "identhousehold.txt"
regionfile = "region.txt"


'open the files : to receive Total Income/Household
labelregionfile = "labelregion.txt"

Rem Open file for reading
identhouseholdfile = "identhousehold.txt"
Open folder & identhouseholdfile For Input As #1
Open folder & regionfile For Input As #2


Rem Open file for export
Open folder & labelregionfile For Output As #3


'rem read the first line and initialize scaleValue
    Input #1, household
    Input #2, region
    
    
    Rem Catch the current household
    householdCurrent = household
    
    Rem initialize the value of the income
    labelregion = region
    
    
    j = 1
    
    
   While Not EOF(1)
     j = j + 1
    Input #1, household
    Input #2, region
     
     If Not (household = householdCurrent) Then
            
    
        Rem The first time we did quit the current household, we may save the current value
        Rem of region as as the region of the household
        Print #3, region
        
            Rem initilize the household
            householdCurrent = household
            End If
    
Wend

        Print #3, region

Close #1, #2, #3
MsgBox "Total Incom found and created in incometotal.txt, and ready to be used"
End Sub
\end{lstlisting}

\newpage
\noindent \textbf{(5b) Creation of the Household Chief Gender label file}.\\

\begin{lstlisting}
Sub createLabelVariableHouseHoldA()

Dim identhouseholdfile As String, household As String
Dim genderfile As String, poswrchieffile As String, labelgenderfile As String, labelgender As String
Dim gender As String, poswrchief As String

Rem Use the right folder where the file are located
folder = "C:/Data_gslo/gslo/bdm/bdm/packagevb6/mali/elim1/"


Rem files needed in text formats : identhousehold , gender, poswrchief
identhouseholdfile = "identhousehold.txt"
genderfile = "gender.txt"
poswrchieffile = "poswrchief.txt"


'open the files : to receive Total Income/Household
labelgenderfile = "labelgender.txt"

Rem Open file for reading
identhouseholdfile = "identhousehold.txt"
Open folder & identhouseholdfile For Input As #1
Open folder & genderfile For Input As #2
Open folder & poswrchieffile For Input As #3


Rem Open file for export

Open folder & labelgenderfile For Output As #4


'rem read the first line and initialize scaleValue
    Input #1, household
    Input #2, gender
    Input #3, poswrchief
    
    
    Rem Catch the current household
    householdCurrent = household
    
    Rem initialize the value of the income
    labelgender = gender
    
    
    j = 1
    
    Rem initialize labelgender to  "XXX"
    labelgender = "XXX"
    Rem test is this individual is household chief. If yes, take the gender as gender of the household chief
    If Trim(poswrchief) = "1" Then
                labelgender = gender
    End If
    
   While Not EOF(1)
     j = j + 1
    Input #1, household
    Input #2, gender
    Input #3, poswrchief
     
     
     If (household = householdCurrent) Then
            
        Rem test is this individual is household chief. If yes, take the gender as gender of the household chief
            If Trim(Str(poswrchief)) = "1" Then
                labelgender = gender
            End If
            
    Else
    
        Rem The first time we did quit the current household, we may save the current value
        Rem of scaleValue as the scale of the the the household which is just completed
        Print #4, labelgender
        
        
        Rem Initialize the sclale and the size for the new household
                
                householdCurrent = household
                
                Rem initialize labelgender to  "XXX"
                labelgender = "XXX"
                    
                    
                    Rem test is this individual is household chief. If yes, take the gender as gender of the household chief
                    If Trim(Str(poswrchief)) = "1" Then
                      labelgender = gender
                     End If
    End If
    
Wend

        Print #4, labelgender

Close #1, #2, #3, #4
MsgBox "gender label created and ready to be used"
End Sub
\end{lstlisting}
\newpage
\noindent \textbf{ (6) Example of categorical income variable : case of ELIM1, Mali}.\

\begin{lstlisting}

Sub transforIncomeMali()

Dim j As Double, newchain As String, valeurIncome As Double, incomefile As String, incomeNTfile As String
Rem Use the right folder where the file are located
folder = "C:/Data_gslo/gslo/bdm/bdm/packagevb6/mali/elim1/"


Rem files needed in text formats : identhousehold , age, gender, poswrchief, sizehousehold, scaleOxford
incomeNTfile = "monthlyincomeNT.txt"
Rem open the files : Income not yet transformg
Open folder & incomeNTfile For Input As #1


Rem files to receive new income
incomefile = "monthlyincome.txt"
Rem open the files : Income not yet transformg
Open folder & incomefile For Output As #2



j = 0
While Not EOF(1)
        Input #1, newchain
        newchain = Trim(newchain)
     j = j + 1
     valeurIncome = 0
     Select Case newchain
            Case "A": valeurIncome = 29000 / 2
            Case "B": valeurIncome = (29000 + 50000) / 2
            Case "C": valeurIncome = (50000 + 100000) / 2
            Case "D": valeurIncome = (100000 + 150000) / 2
            Case "E": valeurIncome = (150000 + 200000) / 2
            Case "F": valeurIncome = (200000 + 30000) / 2
            Case "G": valeurIncome = (300000 + 500000) / 2
            Case "H": valeurIncome = (500000 + 750000) / 2
            Case "U": valeurIncome = (750000 + 1000000) / 2
            Case "J": valeurIncome = (1000000 + 1500000) / 2
            Case "K": valeurIncome = (1500000 + 2500000) / 2
            Case "L": valeurIncome = (2500000 + 3500000) / 2
     End Select
     Print #2, valeurIncome
Wend

Close #1
MsgBox j
'MsgBox "Number of households = " & nbrhouseholds & " Files scalefaofam.txt and sizehousehold.txt ready to be used"
End Sub
\end{lstlisting}

\end{document}